\documentclass{aastex63}
\pdfoutput=1

\newcommand{\rev}{ }
\usepackage{amsmath}
\usepackage{rotating}
\accepted{2021 Jun 9}
\submitjournal{ApJ}

\begin{document}

\title{The Unusual Weak-Line Quasar PG1407+265 and its Foreground $z\sim0.7$ X-ray Cluster}
\shorttitle{PG1407+265 and Cluster}
\shortauthors{McDowell et. al.}

\correspondingauthor{Jonathan McDowell}
\email{jcm@cfa.harvard.edu}
\author[0000-0002-7093-295X]{Jonathan C. McDowell}
\affiliation{Center For Astrophysics | Harvard \& Smithsonian \\
60 Garden St, \\
Cambridge, MA 02138, USA}
\author[0000-0002-0905-7375]{Aneta Siemiginowska}
\affiliation{Center For Astrophysics | Harvard \& Smithsonian \\
60 Garden St, \\
Cambridge, MA 02138, USA}
\author[0000-0002-3993-0745]{Matthew Ashby}
\affiliation{Center For Astrophysics | Harvard \& Smithsonian \\
60 Garden St, \\
Cambridge, MA 02138, USA}
\author[0000-0001-8509-4939]{Katherine Blundell}
\affiliation{Department of Physics, University of Oxford, Keble Rd, Oxford OX1 3RH, UK}
\author{Luigi C. Gallo}
\affiliation{Department of Astronomy \& Physics, Saint Mary's University, 923 Robie Street, Halifax, Nova Scotia, B3H 3C3, Canada\\}

\begin{abstract}
We present new observations of the odd $z=0.96$ weak-line quasar
PG1407+265, and report the discovery of CXOU J140927.9+261813, a $z=0.68$
X-ray cluster.  Archival X-ray photometry spanning nearly four decades  
reveals that  PG1407+265 is variable at the 1\,dex level on a timescale
of years.  V-band variability is present with an amplitude less than 0.1 mag.
The emission-line properties of PG1407+265 also reveal clear
evidence for a powerful inflow or outflow due to near- or
super-Eddington accretion, having a mechanical luminosity of order
$10^{48}$\,erg\,s$^{-1}$.  Our follow-up {\sl Chandra} {\rev exposure centered on
this object reveal } a foreground  
$z=0.68$ cluster roughly $1\arcmin\times1\farcm5$ in extent, offset to
the east of PG1407+265, roughly coincident with the $z=0.68$ radio galaxy
FIRST\,J140927.8+261818.  This non-cool-core cluster contributes about
10\% of the X-ray flux of PG1407+265, has a mass of $(0.6-
5.5)\times10^{14}\,M_\odot$, and an X-ray gas temperature of
{\rev ($2.2-4.3$)\,keV}. Because the projected position of the quasar lies
at about twice that of the cluster’s inferred Einstein radius, lensing
by the cluster is unlikely to explain the quasar's {\rev unusual} properties. We also discuss
the evidence for a second cluster centered on and at the redshift of the quasar.
\end{abstract}


\section{Introduction} \label{sec:intro}

The prototypical X-ray-loud weak-line quasar (WLQ), PG~1407+265, is a
luminous $z\sim1$ AGN with a probable strong outflow (see Section~\ref{section:qso}) but
only a weak, albeit relativistic, radio jet {\rev on parsec scales} \citep{BBB03}. This distinguishes it from
the X-ray weak WLQs studied by \cite{Luo15} and the reason for its
unusual properties - such as weak emission lines and large line velocity
shifts \cite{McDowell95} - remains unresolved.

Luminous AGN are frequently found in BCGs (Brightest Cluster Galaxies) -
the massive elliptical galaxies in the centers of clusters. Such {\rev luminous} quasars, hosting supermassive black
holes (SMBH) with masses $>10^9 M_{\odot}$ (\citealt{McConnell11}, \citealt{McConnell12}),
are believed to be growing via accretion at near-Eddington rates.
Optical studies by \cite{Ellingson91} showed that radio-loud quasars in particular   
are often found in rich clusters.
Cluster formation at $z>1$ must have been influenced by the quasar phase of the BCG, and so
studies of the properties of clusters hosting such quasars provide    
important information about the connection between the BCG and the
cluster halo, the cluster heating and the feedback process, as well as
cluster scaling relations and evolution.
However, there are few known
X-ray-luminous clusters associated with bright quasars at
redshifts low enough for detailed study - e.g. 3C186 at $z = 1.2$ (\citealt{Aneta05},\citealt{Aneta10}), H1821+643 at $z=0.30$ \citep{Russell10},
PKS1229-021 at $z=1.05$ \citep{Russell12}.
Detecting such clusters in the
presence of the bright quasar X-ray emission has proven challenging.
Surveys using the high spatial resolution of {\sl Chandra}
are only now revealing previously missed clusters (e.g. CHiPS, \citealt{Sombo21}).

Here we report the detection of diffuse X-ray emission in the XMM image 
of PG1407+265, together with followup high spatial resolution X-ray observations with {\sl Chandra}
and optical spectroscopy of galaxies in the field. 
We  report the discovery of a foreground ($z=0.68$) X-ray cluster along the line of sight, and describe
what is now known about the unusual quasar PG1407+265.

In this paper we adopt the Planck 2018 cosmology of $H_0 = 67.4~\mbox{km s$^{-1}$ Mpc$^{-1}$}$, $\Omega_\Lambda=0.685$ \citep{Planck18}.

\section{PG1407+265: an extreme, face-on, outflowing AGN } \label{section:qso}

\begin{figure}[ht]

\includegraphics[width=4.5in]{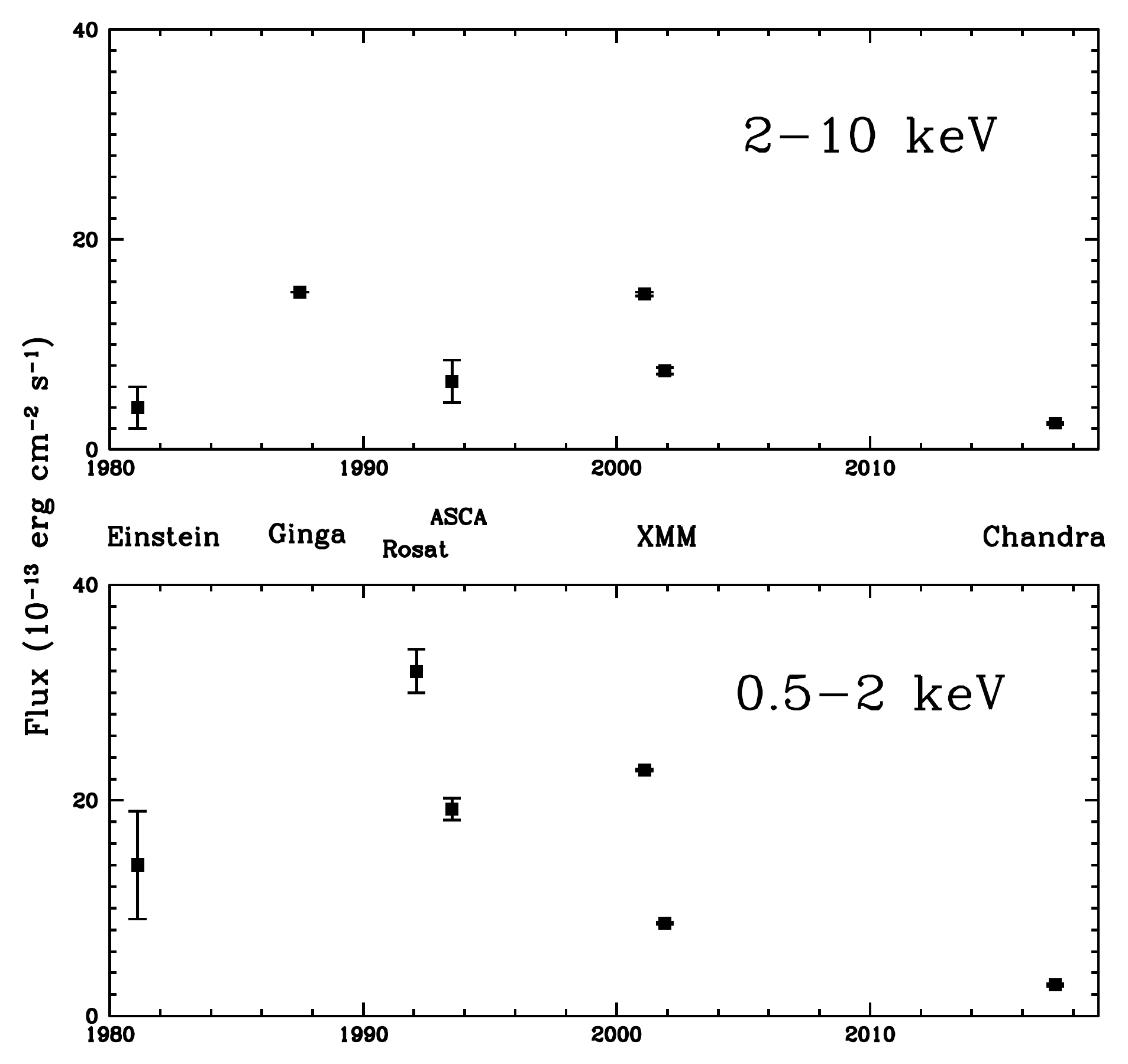}
\caption{
Historical X-ray light curve of PG1407+265. Upper panel: Flux in 2-10 keV energy range.
Lower panel: Flux in 0.5-2 keV energy range. See Table~\ref{table:xflux} for observational details  
and references. \label{fig:lcfig}
}
\end{figure}

Discovered in the
Palomar Green survey \citep{SchmidtGreen83} as a 15th-magnitude object, an extensive
multiwavelength study \citep{McDowell95} confirmed
PG1407+265's classification as a quasar and drew attention to its
unusual combination of emission lines with very low equivalent widths
and large velocity shifts. Specifically, the high-ionization lines are blueshifted up to 13000
km s$^{-1}$ relative to the low-ionization ones, and could indicate a massive
outflow (see e.g. \citealt{Tadhunter08}).  This X-ray-bright and optically bright AGN has
been observed by {\sl Einstein, Rosat, Ginga, ASCA} and {\sl XMM} (Fig.~\ref{fig:lcfig}, Table ~\ref{table:xflux}), but
never previously by {\sl Chandra}. 
Comparison of the historical X-ray observations from
1981 to 2001 {\rev shown in Figure~\ref{fig:lcfig} and detailed in Table~\ref{table:xflux} } indicates 
that the quasar varies by an order of magnitude in X-ray luminosity, well in excess of the
uncertainties of cross-mission
comparison.
The XMM data \citep{Gallo06} revealed factor of 2 variability on several-month
timescales, with high and low spectral states interpreted as being due
to the known jet variability {\rev on parsec scales \citep{BBB03}.}

\begin{table}[ht]
{\small
\begin{tabular}{llllll}
\hline
Date &  Instrument & Seq &    Absorbed Flux & ($10^{-13} \mbox{erg cm$^{-2}$ s$^{-1}$}$)  &Ref\\
     &             &     &    0.5-2 keV & 2-10 keV & \\
\hline
&&&&&\\
1981 Jan  & Einstein IPC & 5381 & 14 $\pm$5&  4?       & \cite{Elvis94}\\  
1987 Jun & Ginga LAC   &     & -        &15  &\cite{Williams92}\\  
1992 Jan &Rosat PSPC-B &RP700359&  32$\pm$ 2   & -        & \cite{McDowell95}\\
1993 Jul & ASCA      & 70024000   & $19\pm 1 $ &$7 \pm 1$ & \cite{ReevesTurner00}\\  
2001 Jan & XMM EPIC  & 0092850101 & 22.8$\pm$0.1 & 14.8$\pm$ 0.2  & \cite{Gallo06}\\
2001 Dec & XMM EPIC  & 0092850501 & 8.6 $\pm$0.1  & 7.5 $\pm$ 0.3 & \cite{Gallo06}\\
2017 Mar & Chandra ACIS&  18265   &2.9 $\pm$ 0.1& 2.4 $\pm$0.1  & This paper\\
\hline
\end{tabular}
}
\caption{X-ray observations of PG1407+265 over 36 years. The {\sl Chandra}~flux is the lowest
level ever observed for this object. Fitted spectral models in the cited
papers are used to recalculate fluxes in the 0.5-2 and 2-10 keV bands when these
were not given explicitly. Flux uncertainties are approximate and take into
account spectral slope uncertainties.\label{table:xflux}}
\end{table}

Optical spectrophotometry \citep{McDowell95} showed that the low-ionization
lines such as H$\alpha$ have a redshift $z =0.96$, while the high-ionization lines such as
CIV have a significantly lower redshift of
$z=0.92$, indicating the presence of a fast outflow or inflow. We tentatively take $z=0.96$ to
represent the system redshift, since the low-ionization lines are unlikely to be part
of the outflow.   Absorption features at $z$=0.575, 0.600, 0.683, and
0.817 have been detected by \cite{Lehner18} using HST COS, and are presumably absorption due to
galaxies along the line of sight.

\begin{figure}[ht]
\includegraphics[width=4.5in]{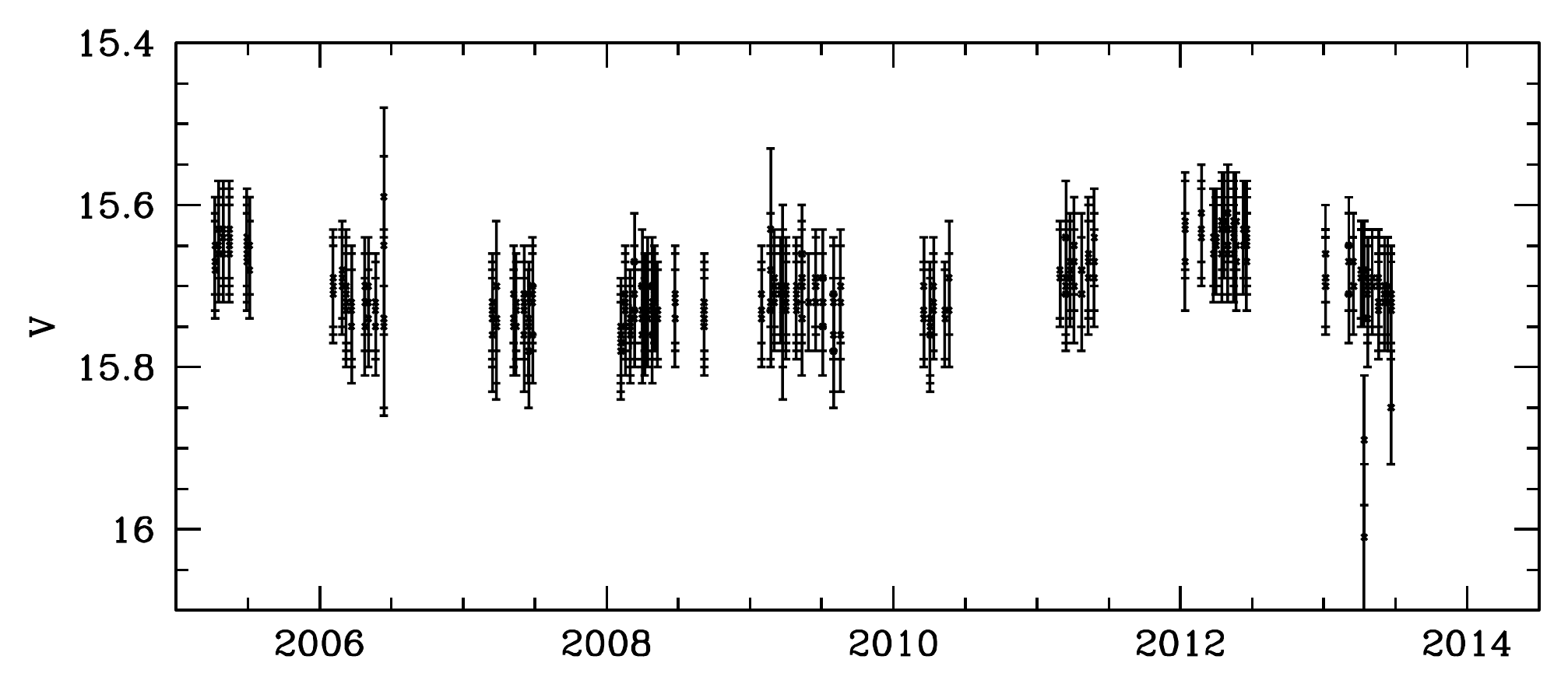}
\caption{ Catalina Sky Survey V magnitudes for PG1407+265.
\label{fig:css}
}
\end{figure}

The Catalina Sky Survey \citep{Catalina}
observed PG1407+265 from 2005 to 2013. The observed (Johnson, Vega-based)
V magnitudes are consistent with the 
measured $V=15.7$ from \cite{SchmidtGreen83}
but indicate a 0.07 mag variability on a 2-year timescale (Fig. \ref{fig:css}).

\begin{figure}[ht]
\includegraphics[width=3.5in]{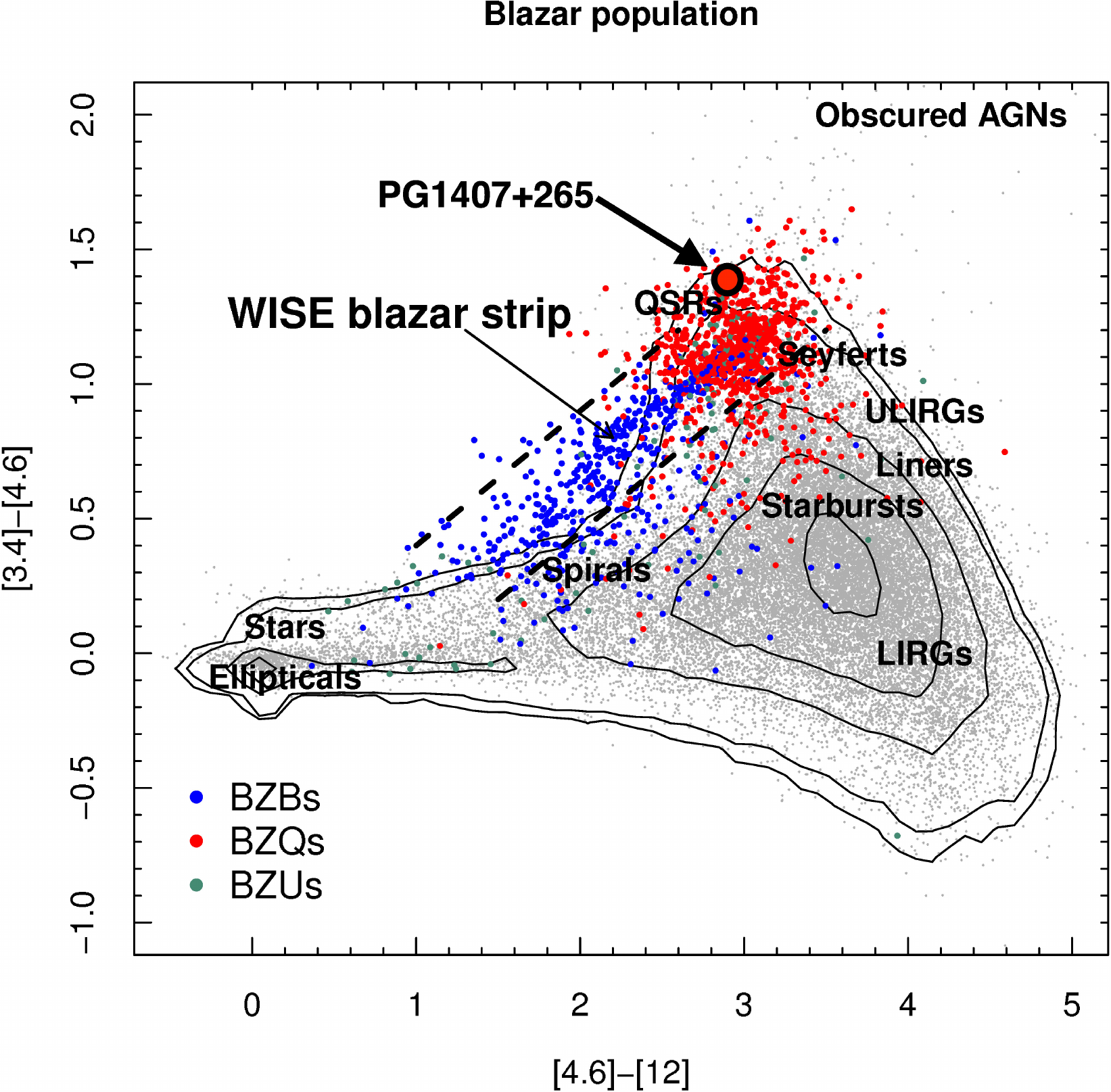}
\caption{
The $[3.4]-[4.6]-[12]$\,$\mu$m {\sl WISE} color-color diagram for
thermal sources and blazars from Fig. 1 of \cite{Massaro11}, amended to indicate
where PG1407+265 lies compared to these populations (large red circle
bordered in black). Units are magnitudes. PG1407+265 has $[3.4]-[4.6]=1.38$\,mag, and
$[4.6]-[12]=2.9$\,mag.  Its mid-infrared colors are similar to those of
quasars and to those of the type BZQ blazars defined by \cite{Massaro09} (i.e. flat spectrum
radio quasars with broad emission lines and blazar characteristics, as opposed to classical BL Lacs).
\label{fig:wise}
}
\end{figure}

The WISE mid-IR colors of the quasar \citep{Cutri13}
are 
[3.4]-[4.6] = 1.38 mag, [4.6]-[12] = 2.00 mag. (The WISE photometry bands
are identified by [$\lambda$]  where $\lambda$ is the effective wavelength in microns.)
Although the object has some blazar-like properties, 
it lies at the edge of the locus for normal quasars discussed in \cite{Massaro11} (Fig. \ref{fig:wise}).

\begin{figure}[ht]
\includegraphics[width=5.5in]{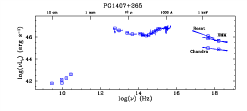}
\caption{Updated spectral energy distribution of the quasar PG1407+265, dereddened
by 0.03 mag and transformed to the rest frame using the Planck cosmology.
Compare with Fig 8. of \citep{Elvis94} which was based on IRAS, IUE and Einstein data.
Data from HST are included (\cite{McDowell95}), as well as from WISE, XMM and Chandra (this paper).
Updated VLA data \citep{Barvainis96} and a IRAS 60 micron measurement \citep{SH09} have been added.
\label{fig:sed}
}
\end{figure}

An updated spectral energy distribution for the quasar is presented in Figure \ref{fig:sed}.
{\rev The optical-to-X-ray spectral index is defined as \citep{Tananbaum79}
\[
\alpha_{ox} = \frac{ \log( L(\nu1) / L(\nu2) )}{ \log( \nu1/\nu2 )}
\]
where $\log \nu1 = 17.684$ corresponding to 2 keV, $\log \nu2 = 15.079$ corresponding 
to $2500$\AA
and $L(\nu)$ is the luminosity per unit frequency in the quasar frame. For PG1407+265
the value of $\alpha_{ox}$ prior to the Chandra observations discussed here was -1.26,
corresponding to a 
luminosity-dependent 
X-ray brightness $\Delta \alpha_{ox}$
relative to typical SDSS quasars \citep{Gibson08} of +0.63, justifying our description
of it as `X-ray-loud' compared to the X-ray-normal WLQs found in \cite{Luo15}.
However, the fainter state seen in 2017 corresponds to $\alpha_{ox}=-1.68$ or $\Delta \alpha_{ox} = +0.21$,
putting the object back in the X-ray-normal range for the time being.
}

The observed bolometric luminosity of the quasar based on the \cite{McDowell95} observations was $3.3\times 10^{47}
\mbox{erg s$^{-1}$} = 5.9
M_\odot c^2 \mbox{yr}^{-1}$ implying a black hole mass of $ M = 2.5 \times 10^9 M_\odot
(L/L_{\mbox{Edd}})^{-1}$. Using the MgII and H$\alpha$ FWHMs and the line
width-continuum luminosity mass scaling relationship of 
\cite{Vester09} we derive a formal black hole mass estimate
for PG1407+265 of $1\times 10^{10}M_\odot$ corresponding to $L/L_{\mbox{Edd}} =
0.25$. However extrapolating the calibration of the relationship to the highest luminosities is uncertain \citep{ShenLiu12}.
\cite{Marcu20}, using an accretion model, estimate a larger mass of $7.9^{+5.1}_{-2.9} \times 10^{9} M_\odot$ and
the true mass is probably at least $5\times10^9
M_\odot$  (in agreement with \citealt{Hryn10});
the velocity shifts suggest the
presence of strong outflows due to near- or super-Eddington accretion
(i.e. with mechanical luminosity $>L_{\mbox{Edd}} \sim (6-12)\times 10^{47} \mbox{erg s$^{-1}$}$).
It may represent
the luminous quasar phase which \cite{YuTrem02}
associate with rapid black hole growth. 

If PG1407+265 were in a large galaxy cluster, the existence of such a wind
would make it a good candidate to exemplify
models such as that of \cite{King09}
where it is proposed that the 
super-Eddington wind from an AGN with a mass in excess of $10^{9} \mbox{M$_\odot$}$ is the source of cluster reheating,
Contrariwise, evidence for a cool core cluster
would suggest that quasar mode heating is unlikely to be widespread. This movitated
the search, discussed below, for an associated cluster of galaxies.

Early quasar surveys relied on the strong emission lines of typical quasars for their discovery.
Weak-line quasars, with CIV equivalent width less than 10\AA, are rare.
One other relatively bright weak-lined object, PHL 1811 \citep{Leighly07}
has been extensively studied, but it is notable for its {\it weak} X-ray
emission. {\rev \cite{Luo15} and \cite{Ni18} 
identify further WLQs and investigated the prevalance of unusually weak X-ray
emission in the group. \cite{Luo15} also describe a class of consistently X-ray-weak
so-called `PHL 1811' analogs (see also \citealt{Shemmer09}, \citealt{Plotkin10}, \citealt{Wu12}).
They present evidence that the latter group  (but not PHL 1811 itself) and about 
half of their WLQ sample are X-ray-weak due to obscuration by outflowing material
that also shields the disk from the central object. If the shielding material
does not block the line of sight, one observes an X-ray-normal WLQ; about half of the \cite{Luo15}
WLQ sample were X-ray normal WLQs.
PG1407+265 is consistent with this scenario as one of 
the X-ray normal WLQs, because it is viewed mostly face-on and 
obscuration is less likely. The other X-ray-normal WLQs, such
as the much less luminous 19th-magnitude SDSS J110938.50+373611.7
(\citealt{Plotkin08},\citealt{Wu12}) at $z$=0.4, are comparatively faint;
PG1407+265 is the nearest examplar of the class and is the only one whose X-ray count rate is enough for detailed
study.   }

The model of PG1407+265 as largely face-on is supported by the multi-epoch radio study of \cite{BBB03}.
They presented evidence suggesting that the object is
an intrinsically radio-quiet quasar amplified by a stunted
pole-on relativistic radio jet. The jet's two main radio knots are separated by a projected
distance of 20 pc (2.5 mas), and there is no evidence for radio jet activity on kpc scales.
The small physical scale of the jet suggests that the object may be related to GPS/CSS sources \citep{ODea20}
and radio-quiet-to-loud transition objects (\citealt{Nyland20}, \citealt{Wolowska21}).

\section{Observations}

\subsection{Reanalysis of the XMM observations}

\vskip 0.1in

\begin{figure}[ht]
\includegraphics[width=3.5in]{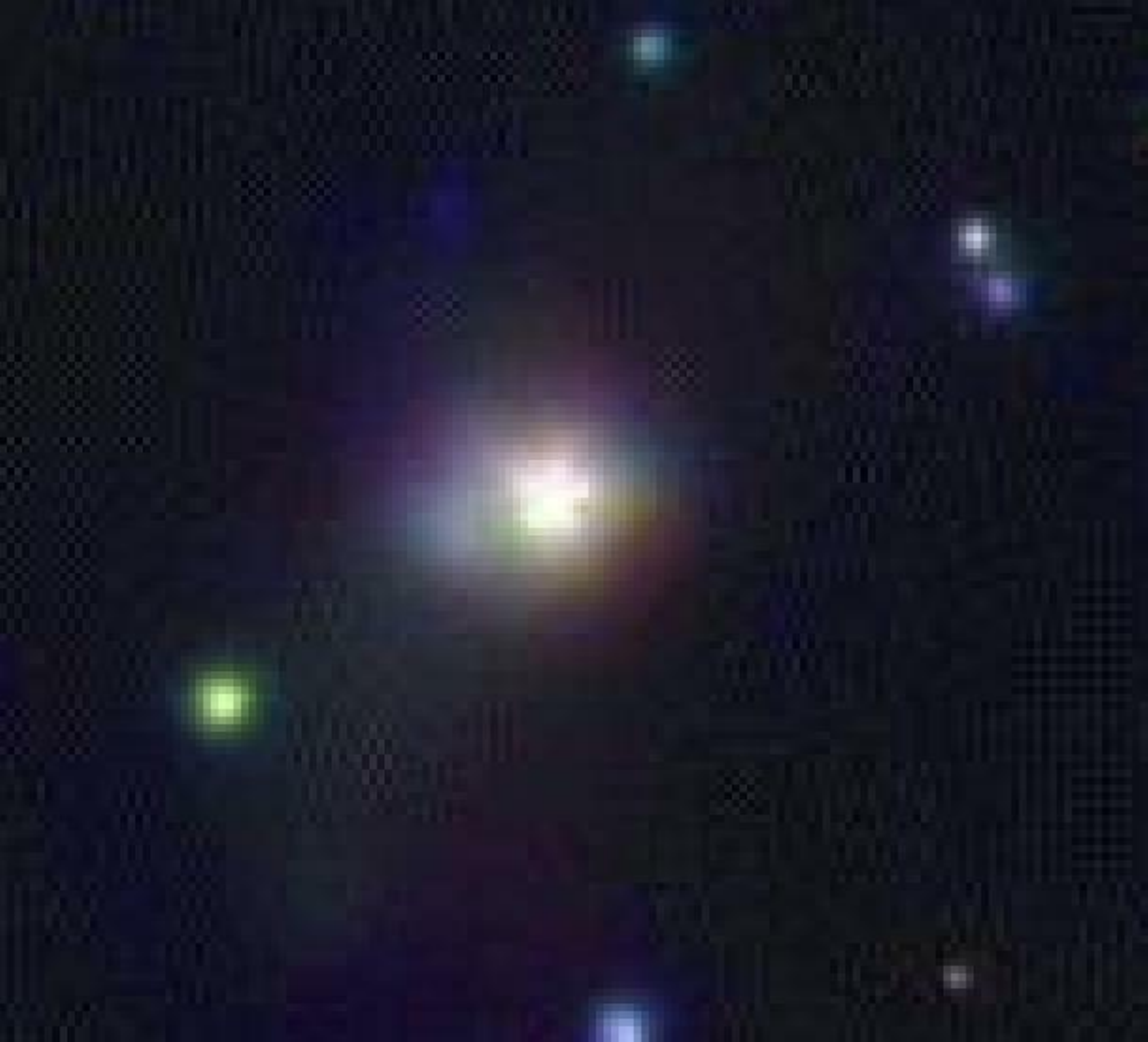}
\includegraphics[width=3.5in]{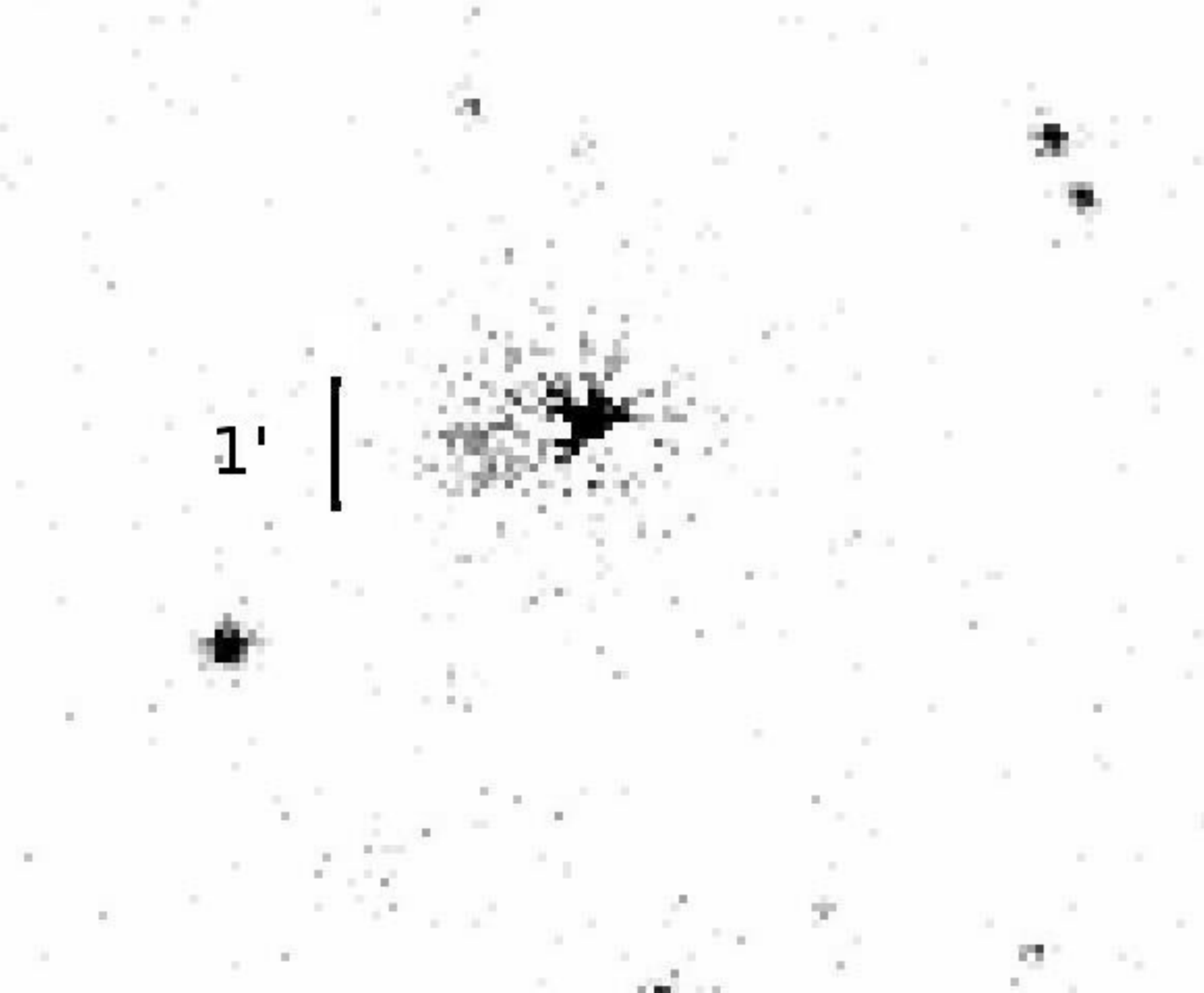}
\caption{
(a) XMM 41 ks MOS1 three-color image, showing extended emission on 1 arcmin
(500 kpc) scales coincident with scale of optical cluster.
(b) XMM 41 ks MOS1 image with partial PSF subtraction,
showing extended emission. The scale bar length is 1 arcmin. North is at top in both images.
\label{fig:xmm}
}
\end{figure}

We reexamined the  2001 XMM observations of the quasar previously reported by \cite{Gallo06}.
The standard SAS archival data products were used \citep{XMM19}.
We used PSFs generated from HR1099 and from PKS0558-504 data and manually scaled to match the central regions
of the PG1407+265 data; the results were similar in both cases. In the second XMM observation the quasar was significantly
fainter than in the first one.
In the XMM MOS1 and MOS2 data taken while PG1407+265 was in this lower state we detect
faint, mildly asymmetric extended emission out to one arcminute radius ($\sim 500$ kpc)
after PSF subtraction
(see Fig. \ref{fig:xmm}).
We interpret this emission as a luminous X-ray cluster.

The Sloan $z$-band image of PG1407+265 shows numerous faint galaxies within the XMM PSF 
and \cite{BBB03} detect radio emission from several, suggesting 
that a cluster can be identified with the extended emission.
The low state X-ray luminosity of PG1407+265 measured by XMM is some $8\times 10^{45} \mbox{erg s$^{-1}$}$ which is comparable
to the most luminous X-ray clusters (e.g. RXJ 1347.5-1145, \cite{Schindler95}),
so it is not unreasonable that cluster emission could contribute significantly
to the total X-ray flux. From the XMM data we estimated
that the cluster X-ray luminosity is at least a few $10^{44} \mbox{erg s$^{-1}$}$ but with large and hard-to-quantify uncertainty,
because the difficulty of removing the large XMM PSF makes a flux determination unreliable. We therefore
carried out {\sl Chandra~}observations to investigate the cluster in more detail.

\vskip 0.1in

\subsection{{\sl Chandra}~observations of the quasar}

On 2017 Mar 14 we made a 41.5 ks observation of the PG1407+265 region using the Chandra
ACIS (Advanced CCD Imaging Spectrometer, \cite{Garmire03}) camera, \dataset[ObsID 18265]{https://doi.org/10.25574/18265}.
The object was placed on chip 7 (S3) and a 1/2 subarray was used
(i.e. reading out the central 512 columns
of the 1024-pixel-wide chip). The resulting 512-pixel-wide strip gives enough area
to ensure a good local background  measurement while being narrow enough
to reduce the pileup of the central AGN \citep{POG}. The data were reduced using
CIAO 4.12 and 4.13 (\citealt{Fruscione06},\citealt{CXC20}).

\begin{figure}[ht]
\includegraphics[width=4.5in]{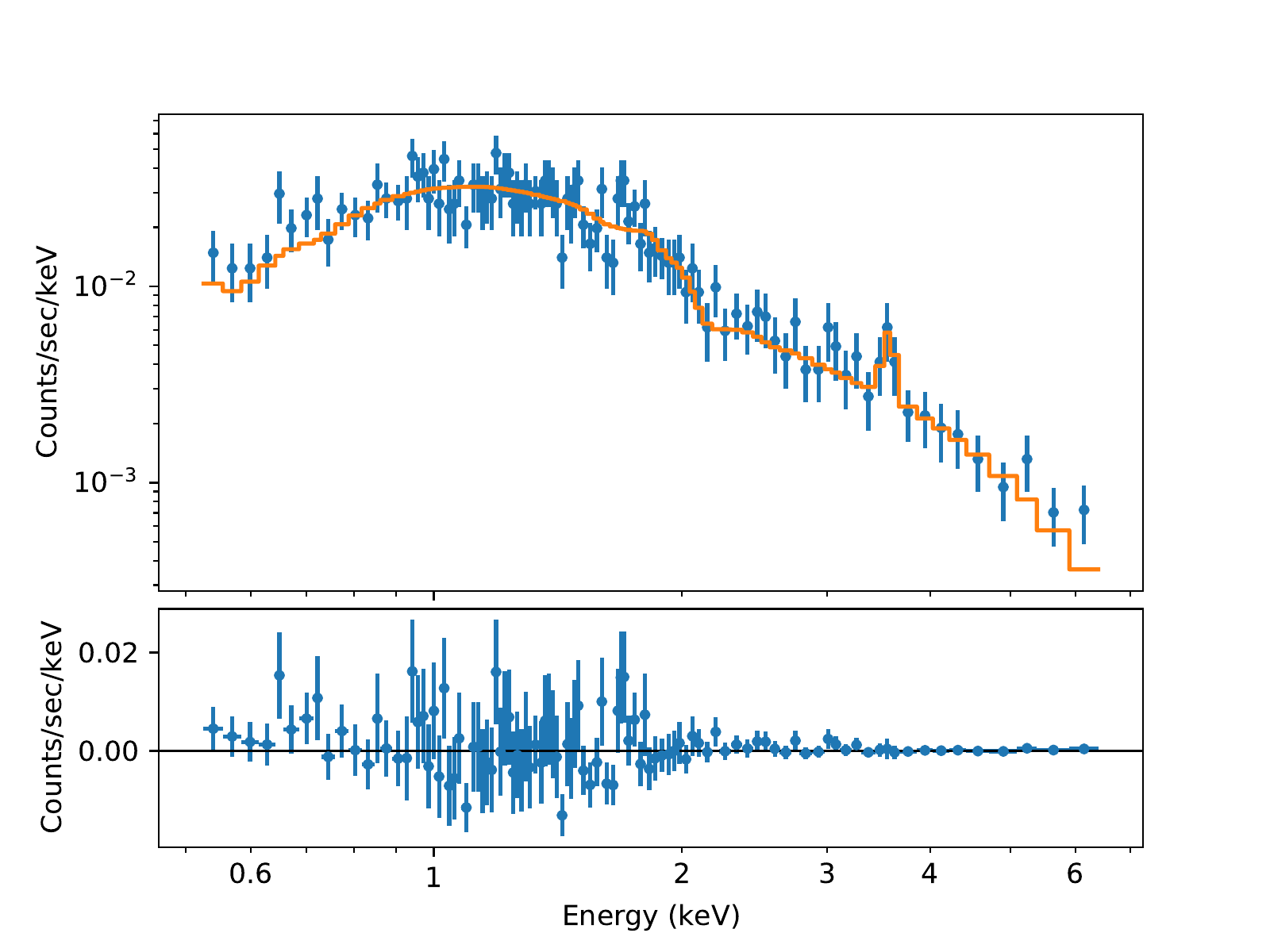}
\caption{
ACIS X-ray spectrum. Absorbed power law fit to the QSO spectrum, with an additional gaussian Fe line,
using the Sherpa fitting package within CIAO. Upper panel: counts and fitted model; lower panel: residuals.
\label{fig:qsp}
}
\end{figure}

The quasar was detected in a low state with 2077 net counts in the broad (0.5-7 keV)
ACIS band. {\rev Using Sherpa \citep{Freeman01}, we fit a power-law model over the 0.5-7 keV range to a PHA (instrumental energy channel) spectrum extracted in a 2-arcsecond
radius circle and grouped to 15 counts per bin, using a chi-squared statistic
and the `levmar' fitting algorithm.} A fixed Galactic absorption of
$1.38\times 10^{20} \mbox{cm$^{-2}$}$ was assumed \citep{McDowell95}.
An energy-dependent aperture correction of less than 10 percent was applied to the effective
area by the standard {\sl specextract}~script.
The fit, shown in Fig. \ref{fig:qsp}, gives a power-law photon {\rev spectral index of $2.26\pm 0.06$
and the unabsorbed flux was found to be
$
  F(0.5-10~\mbox{keV}) = 5.26 \pm 0.48\times 10^{-13} \mbox{erg cm$^{-2}$ s$^{-1}$}
$
(90 percent confidence interval using the {\tt sample\_energy\_flux} routine in {\sl Sherpa}).}
{\rev We checked that fitting instead using the simplex method and the Cash statistic is consistent with these results within the
errors. There is no evidence for intrinsic absorption at the quasar redshift (adding such an absorber
to the fit gives $N_H(\mbox{intrinsic})<10^{21} \mbox{cm$^{-2}$}$).
Restricting the fitting range to rest frame 2-10 keV gives a similar spectral slope of $2.14\pm 0.08$;
on the \cite{Risaliti09} correlation of X-ray spectral slope with Eddington ratio
this would correspond to $L/L_{\mbox{Edd}}\sim 0.4$, consistent
with our suggestion that the object is near-Eddington and with the observation of similarly
high spectral slopes for the `X-ray normal' WLQs studied by \cite{Marlar18}.

The derived flux is a factor of 4 lower than } in the previous faintest state in which
the object has been seen {\rev (in 2001)}. The count rate is fairly stable during the observation, with a 12\% drop from 0.050 to 0.044 counts s$^{-1}$
for the last 10 ks of the exposure.

\begin{figure}[ht]
\includegraphics[width=5.0in]{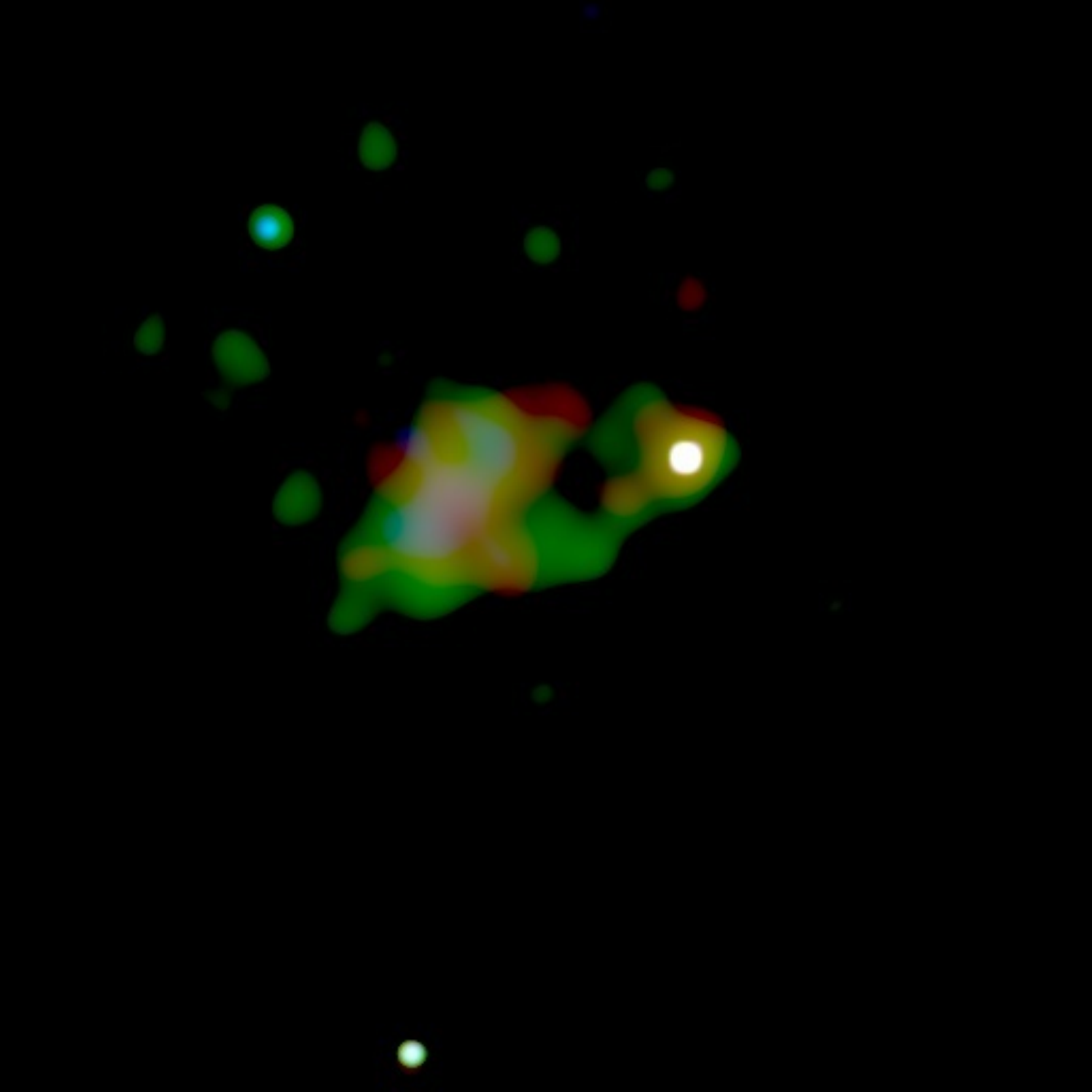}
\caption{
Adaptively smoothed X-ray color image of the {\sl Chandra}~data showing the arcminute-scale diffuse emission.
The standard CSC energy bands are used: Red: 0.3-1.2 keV; Green: 1.2-2.0 keV; Blue: 2.0-7.0 keV.  \label{fig:ximg}
North is at top.
}
\end{figure}

A probable iron line is detected with an observed energy of $3.56 \pm 0.05$
keV and an observed frame equivalent width of 0.14 keV (0.27 keV rest
frame). If this is the {\rev Fe K$\alpha$ line} with rest energy 6.4 keV, the
corresponding redshift is 0.80 which would correspond to a blueshift of
$27000\, \mbox{km s$^{-1}$}$ relative to the inferred systemic velocity, a factor of
two larger than the outflow velocities seen in the UV lines. Alternatively, if
the feature shares the systemic velocity it would correspond to emission
at $6.98 \pm 0.1$ keV. The observed flux of the line is $7.3\times 10^{-15} \mbox{erg cm$^{-2}$ s$^{-1}$}$
and its FWHM is unconstrained.  The line is not seen in the XMM spectra because of 
the higher continuum level during those observations. Refitting the XMM data indicates the line we observe
in Chandra would have needed to be $\sim$5 times brighter to be detectable.

\subsection{The foreground cluster: optical observations}

To the east of the quasar lies a $1\farcm0\times1\farcm5$ region of
diffuse X-ray emission with a total of about 400 net counts.  
The peak of the diffuse X-ray emission is
at 14:09:27.9 +26:18:13 (ICRS), close to (only about 5$^{\prime\prime}$ south
of) the $z=0.68$ radio galaxy FIRST J140927.8+261818.  We interpret the emission
as due to hot gas in a cluster of galaxies associated with the radio galaxy.

By the Chandra naming convention, this cluster's formal name is CXOU J140927.9+261813; however
for ease of reading we will also refer to it in this paper as the East cluster, distinguishing
it from a putative West cluster centered on and assocated with PG1407+265 itself.

\begin{sidewaysfigure}[ht]
\includegraphics[width=9.0in]{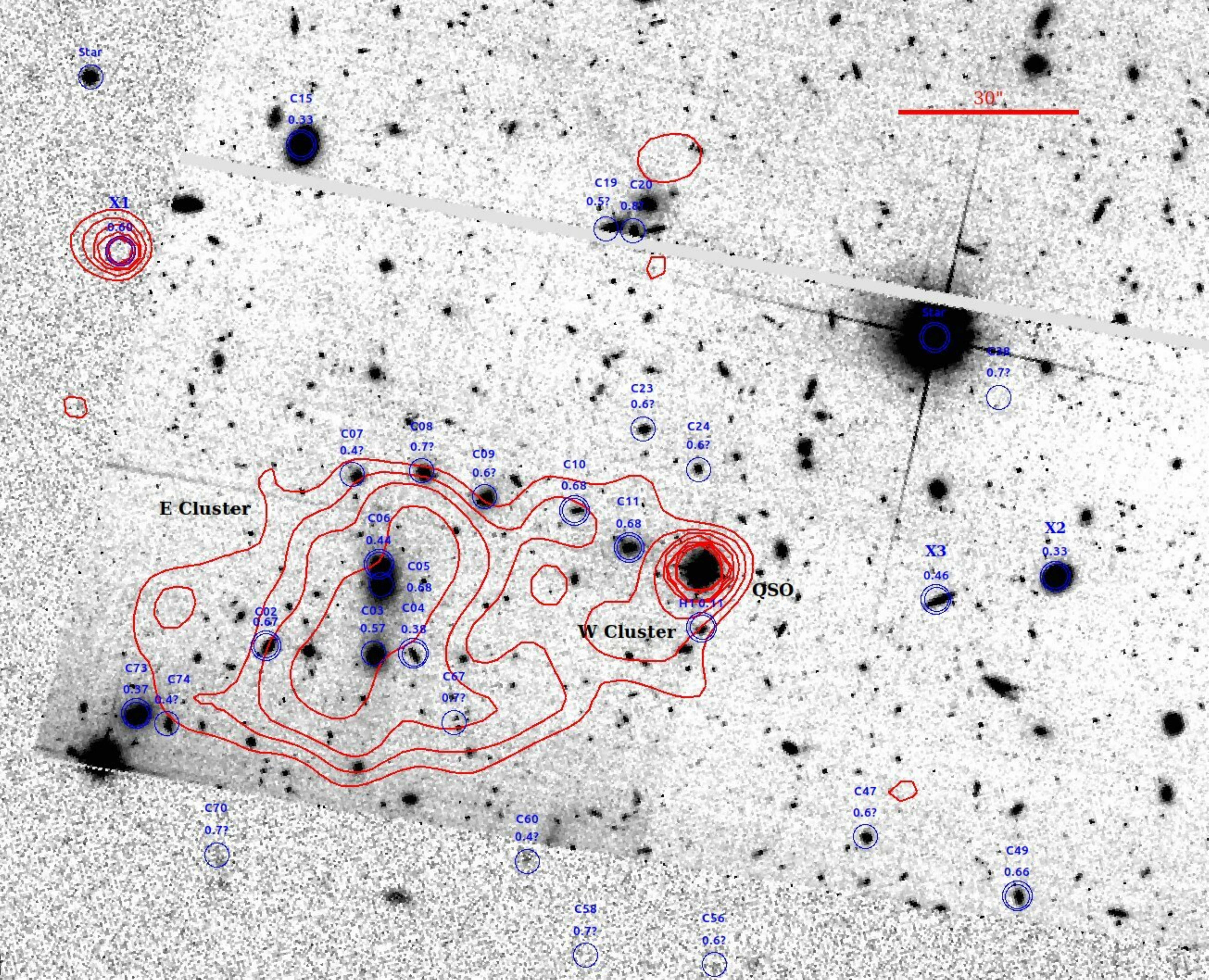}
\caption{
HST ACS image \citep{Lehner15} overlaid on SDSS g-band image,
showing X-ray contours (red) with position of quasar and of the East cluster,
and indicating galaxies with spectroscopic (blue double circles) and photometric (blue single circles) redshifts potentially associated with
the East cluster. Identifiers correspond to those in Table \ref{table:z}.
Scale bar at top right has a length of 30 arcseconds.\label{fig:z}
}
\end{sidewaysfigure}

To understand the galaxy redshift distribution in the field and establish the redshift of the East cluster,
we carried out optical observations in 2019 Apr
with the BINOSPEC fiber spectrograph on the MMT.
Prior to our investigation only the radio galaxy had a published spectroscopic redshift.
There are about 50 further SDSS galaxies of similar optical magnitude
in the field.
SDSS estimates photometric redshifts for most of the remaining objects, but these
have large uncertainties. A number of these photometric redshifts are within two sigma of the
radio source redshift, hinting at possible cluster membership or large scale structure
associated with the cluster. Numerous faint galaxies are visible on an image taken by the Hubble Space Telescope's
Advanced Camera for Surveys (ACS) (ACS-WFC1, F814W filter, sequence \dataset[JCWI06010]{https://doi.org/10.17909/t9-gxy4-t088}, observation date 2016 May 4; 
\cite{Lehner15}; Figure \ref{fig:z}); these could also be cluster members.

For our followup spectroscopic observations BINOSPEC was configured with the 270 lines mm$^{-1}$~grating \citep{Fab19}.
At 1.3 $\mbox{\AA\ pixel}^{-1}$ and a 1 arcsecond wide slitlet, nominal resolution is $R=1340$ corresponding
to about 250 km s$^{-1}$ in the observed frame. Unfortunately as a result of bad weather
we only observed 12 objects, of which four are possibly associated with the radio galaxy.

The BINOSPEC spectra were reduced using the standard BINOSPEC pipeline which extracts wavelength-calibrated
spectra for each slit \citep{Kansky19}. In almost all the
objects, the 4000\AA~break and the Ca H and K lines allow an unambiguous
redshift identification; we then estimated the redshift value by measuring H and K,
and in most cases confirmed the value by also measuring the G band feature (4300\AA), H$\beta$, H$\delta$,
Mg I 5175\AA, and Na D. We estimate the redshift accuracy is of order $\phantom{0}\pm 0.001$.
The three serendipitous {\sl Chandra}~sources, reported here for the first time and denoted X1 to X3 in Table \ref{table:z}, also have narrow OII 3727 and OIII 5007 
emission lines.

Table \ref{table:z}~gives our internal source candidate
identifier, the source name, the photometric redshift from SDSS, and our spectroscopic redshift measurement.
The table also includes SDSS objects whose photometric redshifts are close to 0.7 and thus are possible cluster
members (C05 is the radio source suspected to be the cluster center). A finding chart for the sources overlaid
on the HST ACS image is presented in Figure \ref{fig:z}.
The existence of several objects with $z$ near 0.68 within the X-ray contours, together with the possible $z=0.68$ Fe line in the
cluster X-ray spectrum and the known $z=0.68$ UV absorption feature in the quasar, lead us to propose that
the redshift of the East cluster CXOU J140927.9+261813 is indeed $z=0.68$.

\begin{table}[ht]

\begin{tabular}{llll}
ID  & Name                 & $z_{\rm p}$  & $z$    \\
\hline
\multicolumn{3}{c}{Chandra sources}\\
\hline
X1  & CXOU J140931.0+261913 & &0.599\\
X2  & CXOU J140919.5+261820 & &0.460\\
X3  & CXOU J140921.0+261816 & &0.327\\
\hline
\multicolumn{3}{c}{Candidate cluster members}\\
\hline
C02 & SDSS J140929.25+261808.5 & 0.63 & 0.671\\
C03 & SDSS J140927.93+261807.3 & 0.57 &  - \\
C05 & SDSS J140927.84+261818.7 & 0.45 & 0.68 (SDSS)\\
C08 & SDSS J140927.33+261837.5 & 0.71 &  - \\
C09 & SDSS J140926.56+261833.2 & 0.61 & - \\
C10 & SDSS J140925.45+261831.1 & 0.62 & 0.680\\
C11 & SDSS J140924.78+261824.8 & 0.78 & 0.681\\
C19 & SDSS J140925.06+261917.6 & 0.53 & -\\
C20 & SDSS J140924.73+261917.4 & 0.77 & -\\
C23 & SDSS J140924.61+261844.5 & 0.62  & -\\
C24  & SDSS J140923.92+2618378 &0.56 & -\\
C38 & SDSS J140920.22+261849.7 & 0.70 & -\\
C47 & SDSS J140921.86+261736.9& 0.59 & -\\
C49 & SDSS J140919.99+261727.0 & 0.57 & 0.658\\
C51 & SDSS J140919.94+261647.6 & 0.56 & - \\
C56 & SDSS J140923.73+261715.7 & 0.56 & -\\
C58 & SDSS J140925.31+261717.2 & 0.72 & -\\
C67 & SDSS J140926.93+261755.8 & 0.71 & -\\
C70 & SDSS J140929.86+261733.9 & 0.66 & -\\
C75 & SDSS J140931.05+261913.9 & 0.74 & -\\
\hline
\multicolumn{3}{c}{Other sources in field}\\
\hline
C04 & SDSS J140927.43+261807.2 &      & 0.375 \\
C06 & SDSS J140927.86+261822.2 & 0.47  & 0.435\\
C15 & SDSS J140928.82+261931.6 & 0.29 &  0.328\\
C73 & SDSS J140930.85+261755.6 & 0.40 & 0.374\\
H1  & Anon 140923.9+261812  &  & 0.114\\
\hline
\end{tabular}
\caption{Estimated redshifts of sources in the PG1407+265 field. The ID is an
internal source identifier; $z_{\rm p}$ is the SDSS photometric redshift and $z$ is the
spectroscopic redshift (from our MMT observations except where noted).
We identify object X1 with optical source SDSS J140931.05+261913.9.
\label{table:z}}
\end{table}

\subsection{{\sl Chandra}~observations of the East cluster: spectrum and radial profile}

We filtered the X-ray event list into three energy bands (the standard Chandra Source
Catalog bands of 0.3-1.2\,keV,
1.2--2.0\,keV, and 2.0--7.0\,keV, \citealt{Evans10}) to look for spatial variations of the X-ray spectrum.
The energy boundaries were chosen to give reasonable signal-to-noise in each band.
We then adaptively smoothed each of the
three bands separately with the CIAO {\sl csmooth} tool.  The resulting
three-band background-subtracted X-ray color image of the field is shown
in Fig.~\ref{fig:ximg}; the apparent spectral variations are not statistically significant.

\begin{figure}[ht]
\includegraphics[width=6.0in]{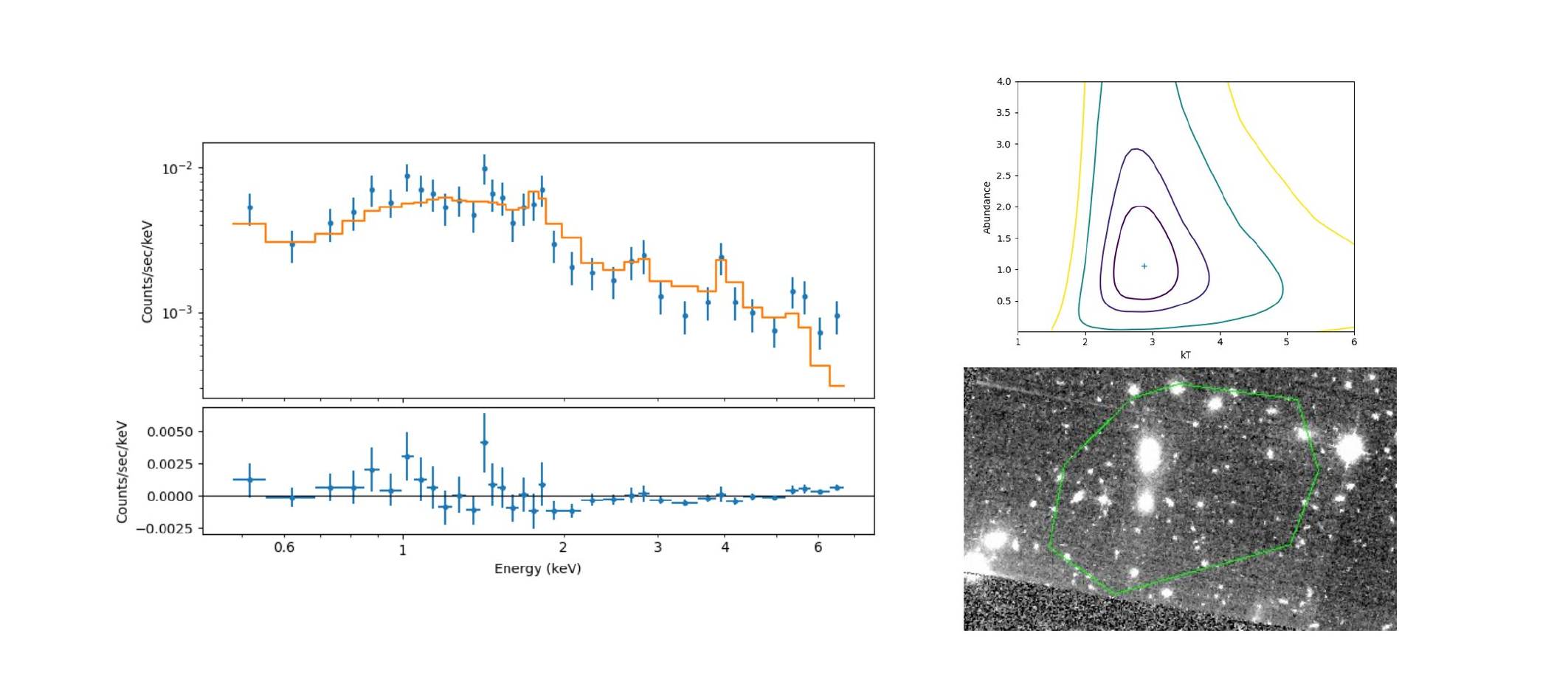}
\caption{
Left panel: APEC fit to the nebular {\rev (East cluster)} spectrum assuming $z=0.68$, showing a redshifted iron line and
two other possible features. The best fit temperature is poorly constrained, 
$kT = 2.9^{+1.6}_{-0.9}$ keV.
{\rev Right upper panel: 0.5, 1, 2 and 3 sigma reduced $\chi^2$ contours (relative to best fit)}
versus temperature and abundance for APEC model spectral fit.
Right lower panel: ACIS image of region with extraction region overlaid (green polygon).
\label{fig:nebsp}
}
\end{figure}

We also extracted a spectrum from the event file in a polygonal region surrounding the East cluster diffuse emission.
The spectrum has 345 net counts within this extraction region in the 0.5-7 keV band. 
{\rev A 5-arcsecond-radius region around the quasar was excluded; no other point sources were detected
within the diffuse emission region. }
We fit an APEC (Astrophysical Plasma Emission Code,
\citealt{Smith01}) model
with solar abundances, a redshift of 0.68, and the fixed Galactic absorption (Fig. \ref{fig:nebsp}). 
{\rev Due to the low number of counts, we performed a simplex fit using the Cash statistic; the
background was fit separately and modelled as a power law with an additional line at 1.75 keV.
}

{\rev The source's temperature is weakly constrained at $2.9^{+1.6}_{-0.9}$ keV, }
and allowing the abundance to vary
does not change the overall fit significantly. 
However, an emission feature is visible at 4.0 keV which is well fit by the redshifted 6.7 keV iron
line produced by the model, as long as the metallicity is around solar or greater. We note this metallicity is larger than
the typical metallicity for non-cool-core clusters of 0.2 reported by \cite{AllenFabian98}.
The line suggests that the cluster may indeed be at the redshift of the radio galaxy (see further, below).
Adding two other lines at observed energies of 2.77 and 5.49 keV improves the fit somewhat. At the radio galaxy redshift
these would correspond to 4.65 keV and 9.22 keV respectively; we are unable to propose interpretation of these features.
Using the MEKAL (Mewe-Kaastra-Leidahl, \citealt{Mewe85}) model instead of APEC gives a very similar fit.

{\rev The unabsorbed flux  of the cluster
(i.e., the flux corrected for Galactic foreground absorption)
 is $F(0.5-2 \mbox{keV}) = (3.6\pm 0.2)\times 10^{-14} \mbox{erg cm$^{-2}$ s$^{-1}$ } $
and
$F(2-10 \mbox{keV}) = (6.5 \pm 1.3)\times 10^{-14} \mbox{erg cm$^{-2}$ s$^{-1}$ } $
(90 percent confidence intervals, using the {\rev Sherpa} {\sl sample\_energy\_flux} command).

If the cluster is at the radio galaxy distance, $L(0.1-2.4 \mbox{keV}) = (6.8\pm 0.6) \times 10^{43} \mbox{erg s$^{-1}$}$.
}
Estimates of the flux and luminosity of the quasar and cluster
are given in Table \ref{tab:xfit}.

Using the temperature-mass relationship derived by \cite{Finoguenov01} for objects with temperatures above 3 keV, we infer
a cluster mass $\log(M/M_\odot) = 14.3^{+0.4}_{-0.5}$ within the conventional $r_{500}$ radius
at which the density drops below 500 times the
cosmological critical density at $z=0.68$.

{\rev
\begin{table}[ht]
\begin{tabular}{llll}
     &  Energy band &  Quasar & East Cluster\\
     &    (keV)      &       &          \\
\hline
Assumed redshift $z$    && 0.96   & 0.68 \\
Unabsorbed flux ($10^{-13} \mbox{erg cm$^{-2}$ s$^{-1}$}$) & 0.5-2 (obs)&$2.88\pm0.18$ & $0.36 \pm 0.02$\\
Unabsorbed flux ($10^{-13} \mbox{erg cm$^{-2}$ s$^{-1}$}$) & 2-10  (obs)&$2.38\pm0.37$ & $0.65 \pm 0.13$ \\
Luminosity ($10^{44} \mbox{erg s$^{-1}$}$)                          & 0.1-2.4 (rest)&$48.4\pm 2.8$&$0.68 \pm 0.06$\\
Luminosity ($10^{44} \mbox{erg s$^{-1}$}$)                          & 2-10    (rest)& $14.0\pm3.8$&$1.4 \pm 0.2$\\
Luminosity ($10^{44} \mbox{erg s$^{-1}$}$)                          & 0.1-10    (rest)&$60.6 \pm4.3$& $2.0\pm 0.2$\\
\hline
\end{tabular}
\caption{Fitted fluxes and luminosities for the quasar and cluster in soft and hard bands.
Values and uncertainties are medians and 90\% confidence intervals derived from the {\sl Sherpa} {\tt sample\_energy\_flux} routine.
Luminosities for the quasar at rest frame energies below 1 keV are provided
for comparison with other sources but are an extrapolation of the fit and therefore not reliable.
Luminosities for the cluster have uncertainties of 8\%, with roughly equal contribution from 
the Poisson error (low number of counts) and the fitting uncertainty in the temperature.
\label{tab:xfit}}
\end{table}

}

\begin{figure}[ht]
\includegraphics[width=3.5in]{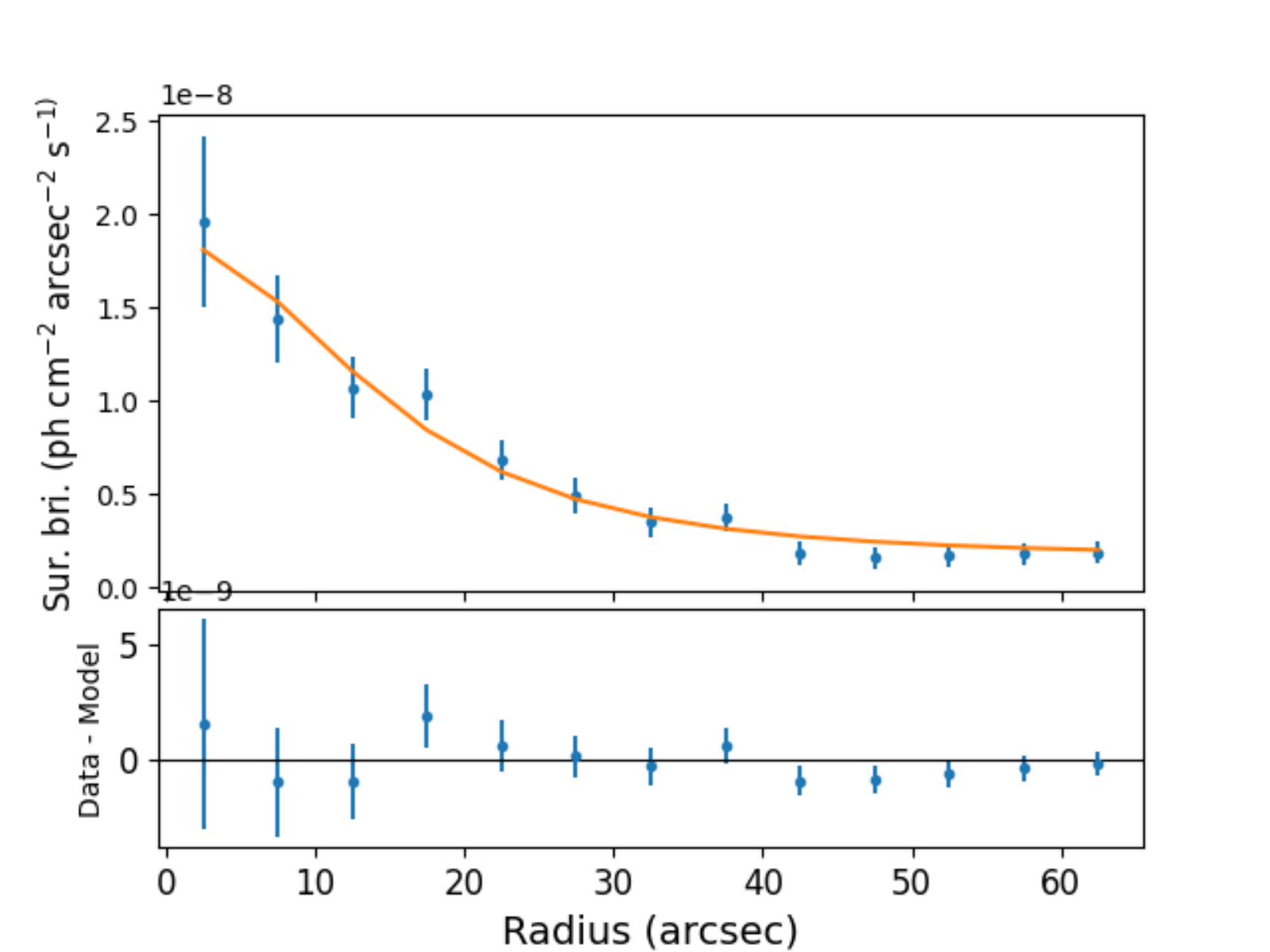}
\includegraphics[width=3.5in]{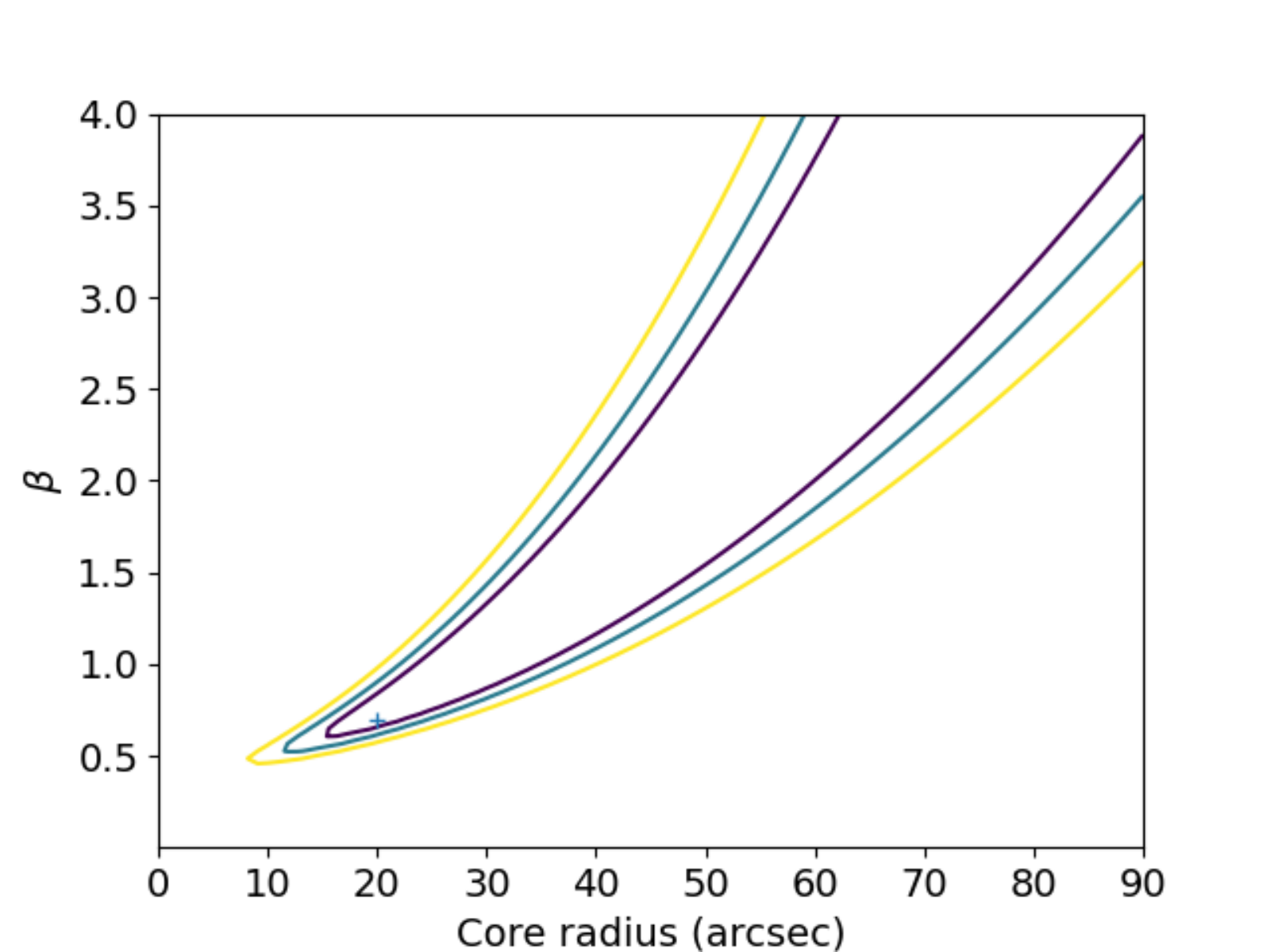}
\caption{
Left: Cluster radial profile and fit. The $x$ axis gives the radius from the
X-ray peak in arcseconds; the $y$ axis is the 0.5--7.0 keV surface brightness in $10^{-8} \mbox{photon cm$^{-2}$ arcsec$^{-2}$ s$^{-1}$}$.
Residuals are shown in the lower left panel. 
\\
Right: 1, 2 and 3 sigma contours of reduced $\chi^2$ for the 
cluster radial profile fit, as a function of core radius in arcseconds ($x$ axis) and beta ($y$ axis).
The parameters are weakly constrained.
\label{fig:xpro}
}
\end{figure}

We also extracted a radial X-ray surface brightness profile for the East
cluster.  An exposure-corrected image was created in the 0.5--7 keV band using
the CIAO {\sl fluximage} script. To prevent contamination of the profile by the bright quasar
itself, a circular region of 10$^{\prime\prime}$ radius centered on
PG1407+265 
was first replaced by photons sampled from the 
same radial distance from the cluster center using the CIAO tool {\sl dmfilth}.
All the regions were visually
inspected to verify that they were not contaminated by bright sources,
and that they did not extend outside the bounds of the exposed areas of
the ACIS detectors.  
The exposure-corrected cluster X-ray
surface brightness in the 0.5--7 keV band was then measured in concentric $5^{\prime\prime}$-wide annuli centered on
the peak of the diffuse emission using the
CIAO tool {\sl dmextract}. A background region $1\farcm5$ away was used to subtract a constant component.
The resulting
background-subtracted radial surface brightness profile for CXOU
J140927.9+261813 is shown in Fig.~\ref{fig:xpro}.

The X-ray surface brightness profile for CXOU J140927.9+261813 was fit
with a beta model using {\sl Sherpa}. Formally the profile is poorly constrained
by the global fit (Fig \ref{fig:xpro}, right panel), but examination of residuals for an ensemble of fits suggests that a core radius of 15 to 35 arcsec
and a beta value of 0.6 to 1.0 provide a reasonable representation of the data. A representative
fit with core radius 20 arcsec and $\beta=0.70$ is
overplotted on the measurements in Fig.~\ref{fig:xpro}.  
Using an isothermal hydrostatic equilibrium model, the mass within 500 kpc derived from this fit is found to be
$\log(M(500 \mbox{kpc})/M_\odot) = 14.2\pm 0.4$. This is consistent with the simple mass-temperature relation
used earlier although we note the 500 kpc radius used is different from that relation's $r_{500}$ radius, whose value is poorly constrained in our data.
The amplitude of the fitted profile is, however, well-constrained:
$
S_0=1.8\pm0.2\times10^{-8} \mbox{ photons cm$^{-2}$ s$^{-1}$ arcsec$^{-2}$}
$
and there is no evidence of a cuspy central peak, so we infer that the cluster is a non-cool-core one.

\subsection{Cluster Lensing}

\cite{McDowell95} raised the possibility that the unusual nature of PG1407+265 might in part be
due to lensing by a foreground object. 
The discovery of exactly such a foreground object 
prompted us to re-examine this possibility in light of the newly available {\sl Chandra} observations.

Our derived East cluster mass of $\log(M/M_\odot) = 14.3^{+0.4}_{-0.5}$ corresponds to an Einstein radius of 0.1 to 0.3 arcmin, compared
to the cluster-center-to-quasar distance of about 1 arcmin. This argues against significant lensing of the quasar
PG1407+265 by the cluster CXOU J140927.9+261813, but deeper observations will be needed to conclude this with confidence.

\subsection{The West cluster: radial profile analysis}

The diffuse emission near the quasar could be foreground emission from the East cluster, in which case
this cluster has a disturbed morphology and is far from virialized - possibly, even, two clusters merging. 
We note that the dark `bay' between
the main East cluster and the emission near the quasar is significant (in a 7 arcsecond circle, total
broad band counts are only 24 compared to 40 in adjoining areas of equal size).

\begin{figure}[ht]
\includegraphics[width=4.0in]{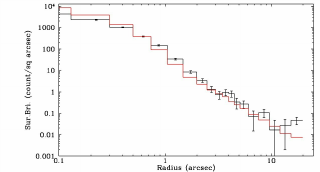}
\caption{
Broad-band (0.5-7 keV) X-ray radial profile centered on the quasar. Black line: log of extracted surface
brightness versus log radial distance from center. Red line: Chart/MARX PSF radial profile with
normalization matched at 1 pixel (0.5 arcseconds). A possible excess near 1-2 arcseconds
is within the uncertainties of the PSF normalization.
\label{fig:qpro}
}
\end{figure}

An alternative possibility is that the emission near the quasar is a second cluster - we will call
it the West cluster - centered on the quasar and at its redshift. We do not have enough counts
in the western emission to detect the putative Fe line in that region. We therefore searched
for evidence of a peak in the extended emission around the quasar point source, but the
current data do not give us a definitive answer: there
is no clear evidence of extent in the quasar image at the few arcsecond
scale. In Fig~\ref{fig:qpro} we show the radial profile of the 0.5-7 keV
count surface brightness distribution centered on the quasar and compare
it with a point spread function made using the ChaRT \citep{Carter03}
and MARX \citep{Davis12} PSF simulation tools. The PSF shown is the average of five
realizations and is normalized by matching the data at a radius of 0.5 arcseconds.
It can be seen that the PSF is slightly narrower than the data, and there is some
suggestion of excess flux near 1 to 2 arcseconds, but we conclude that there is no
definite evidence of extent except beyond 10 arcseconds. Nevertheless, the statistical errors
would allow the presence of a diffuse source with a luminosity of order $10^{44} \mbox{erg s$^{-1}$}$.

A deeper exposure will be needed to decide whether the western emission is part of the East cluster
or is from a West cluster associated with PG1407+265 itself.

\section{Conclusion}

A quarter-century after the study of \cite{McDowell95} the PG1407+265 system remains
rather enigmatic but some aspects of it have come into focus. The face-on
but stunted radio jet may indicate that the quasar is a failed blazar
(\citealt{BBB03}, \citealt{McDowell95})
with a broader rapid outflow explaining the extreme line shifts. 
{\rev These shifts seen at UV and, possibly, X-ray wavelengths
could even be due to interaction with the weak but relativistic radio jet; future work
is needed to explore this possibility.}
The almost-line-of-sight galaxy cluster CXOU J140927.9+261813 complicates
the picture but the observations suggest that we cannot explain away the low equivalent
width lines as being due to lensing. 
The existence of X-ray emission from a second
cluster at the redshift of the quasar is ambiguous. The previous absence of these
clusters from cluster catalogs due to the glare of the quasar emphasizes
the importance of high spatial resolution X-ray imaging of the kind
provided by {\sl Chandra}~and proposed for {\sl Lynx}.

\section{Acknowledgements}

This work was supported by Chandra grant GO6-17117X 
and by the NASA Chandra X-ray Center,
which is operated by the Smithsonian Astrophysical Observatory for and on behalf of the National Aeronautics Space Administration under contract NAS8-03060.
We thank MMTO/TDC staff for making the MMT observations and processing
the data through the standard pipeline.
We also thank R. Saxton for advice on the XMM PSF and D. Burke and B. Wilkes for helpful comments.
We acknowledge the use of the software packages SAS \citep{XMM19}, CIAO (\citealt{Fruscione06},\citealt{CXC20}), Sherpa \citep{Freeman01} and DS9 \citep{Joye03},
as well as Ned Wright's cosmology calculator \citep{Wright06}.
This research has made use of the 
NASA/IPAC Infrared Science Archive, which is operated by the Jet Propulsion Laboratory, California Institute of Technology, under contract with the National Aeronautics and Space Administration;
the SIMBAD database,
operated at CDS, Strasbourg, France, and NASA's Astrophysics Data System Bibliographic Services.
This publication makes use of data products from the Wide-field Infrared
Survey Explorer, which is a joint project of the University of
California, Los Angeles, and the Jet Propulsion Laboratory/California
Institute of Technology, and NEOWISE, which is a project of the Jet
Propulsion Laboratory/California Institute of Technology. WISE and
NEOWISE are funded by the National Aeronautics and Space Administration.
The Catalina photometry is courtesy of the CSS survey,  funded by the National Aeronautics and Space
Administration under Grant No. NNG05GF22G issued through the Science
Mission Directorate Near-Earth Objects Observations Program.

We thank the anonymous referee for their helpful comments on the initial version of this paper.

\par

This is the Accepted Manuscript version of an article accepted for
publication in the Astrophysical Journal.  IOP Publishing Ltd is not responsible
for any errors or omissions in this version of the manuscript or any
version derived from it.  The Version of Record will be available online 
at https://iopscience.iop.org/journal/0004-637X

\clearpage

\bibliography{ms}{}
\bibliographystyle{aasjournal}

\end{document}